\documentstyle[12pt,epsf]{article}

\def\gsim{\raisebox{-.4ex}{\rlap{$\sim$}} \raisebox{.4ex}{$>$}}
\begin{document}
\begin{flushright}
\mbox{FERMILAB-Conf-00/278-T}\\
\mbox{October 2000} \\[0.75in]
\end{flushright}
\begin{center}
        {\large \bf Neutrino Oscillation Scenarios and GUT Model 
	Predictions\footnote{Talk given at the Joint U.S./Japan Workshop 
	on New Initiatives in Lepton Flavor Violation and Neutrino 
	\hspace*{0.23in} Oscillations with Very Intense Muon and Neutrino 
	Sources (Honolulu, HI), October 2-6, 2000.}}
        \\[1.5in]

Carl H. Albright\footnote{E-mail: albright@fnal.gov}\\ 
Fermi National Accelerator Laboratory \\
P.O. Box 500, Batavia, IL 60510 \\[1.5in]
\end{center}

\begin{abstract}

The present experimental situation regarding neutrino oscillations is 
first summarized, followed by an overview of selected grand unified models
which have been proposed to explain the various scenarios with three
active neutrinos and their right-handed counterparts.  Special attention 
is given to the general features of the models and their ability to favor
some scenarios over others.

\thispagestyle{empty}
\end{abstract}
\newpage

\section{Three-Active-Neutrino Oscillation Scenarios}
\subsection{Atmospheric Neutrinos}\label{subsec:atm}
Recent results from the Super-Kamiokande Collaboration~\cite{atm} involving 
atmospheric neutrinos convincingly favor muon-neutrinos oscillating into 
tau-neutrinos rather than into light sterile neutrinos.  The latter possibility
is ruled out at the 99\% confidence level.  In terms of the oscillation
parameters, $\Delta m^2_{ij} \equiv m^2_i - m^2_j$ and $\sin^2 2\theta_{atm}$,
the best fit values obtained are 
\begin{equation}
\begin{array}{rcl}
	\Delta m^2_{32} & = & 3.2 \times 10^{-3}\ {\rm eV^2},\\[4pt]
	\sin^2 2\theta_{23} & = & 1.000,
	\\ & &
\end{array}\label{eq:atm}
\end{equation}
with the latter related to the neutrino mixing matrix elements by 
$\sin^2 2\theta_{atm} = 4|U_{\mu 3}|^2 |U_{\tau 3}|^2$.

\subsection{Solar Neutrinos}
The situation regarding solar neutrinos is considerably less certain.  The 
recent analysis~\cite{SKsolar} by the Super-Kamiokande Collaboration 
involving their 1117 day sample together with the data from the 
Chlorine~\cite{chlor} and Gallium~\cite{gall} experiments favor the large 
mixing angle MSW~\cite{MSW} solution (LMA) and possibly the LOW solution over 
the small mixing angle (SMA) and vacuum (VAC) solutions, with the latter two 
being ruled out at the 95\% confidence level.  Several theory groups analyzing 
the same data suggest instead that while the LMA solution is favored, the other
solutions are still viable at the 95\% c.l.  In fact, a continuum solution
-- the quasi-vacuum solution (QVO) -- stretches between the LOW and VAC regions
with $\tan^2 \theta_{sol} \gsim 1.0$.  The best fit points in the various 
parameter regions found in a recent analysis by Gonzalez-Garcia
and Pe\~{n}a-Garay~\cite{ggpg} are given by
$$\begin{array}{ll}
	SMA: & \Delta m^2_{21} = 5.0 \times 10^{-6}\ {\rm eV^2},\\[2pt]
	     & \sin^2 2\theta_{12} = 0.0024,\\[2pt]
	     & \tan^2 \theta_{12} = 0.0006,\\[4pt]
	LMA: & \Delta m^2_{21} = 3.2 \times 10^{-5}\ {\rm eV^2},\\[2pt]
	     & \sin^2 2\theta_{12} = 0.75,\\[2pt]
	     & \tan^2 \theta_{12} = 0.33,\\[4pt]
\end{array}$$
\begin{equation}
\begin{array}{ll}
	LOW: & \Delta m^2_{21} = 1.0 \times 10^{-7}\ {\rm eV^2},\\[2pt]
	     & \sin^2 2\theta_{12} = 0.96,\\[2pt]
	     & \tan^2 \theta_{12} = 0.67,\\[4pt]
	QVO: & \Delta m^2_{21} = 8.6 \times 10^{-10}\ {\rm eV^2},\\[2pt]
	     & \sin^2 2\theta_{12} = 0.96,\\[2pt]
	     & \tan^2 \theta_{12} = 1.5,\\[4pt]
\end{array}\label{eq:sol}
\end{equation}
Note that $\theta_{12}$ is in the second octant or ``dark side'' for the 
quasi-vacuum region.  An even more recent analysis,~\cite{ggmpgv} which includes
the CHOOZ reactor constraint,~\cite{CHOOZ} modifies the above numbers 
slightly, and sets $\tan^2 \theta_{13} = 0.005$. 

\subsection{Maximal and Bimaximal Mixings}
The Maki-Nakagawa-Sakata (MNS) neutrino mixing matrix, analogous to the CKM 
mixing matrix, can be written as
\begin{equation}
  U_{MNS} = \left(\matrix{c_{12}c_{13} & s_{12}c_{13} & s_{13}e^{-i\delta}\cr
	-s_{12}c_{23} - c_{12}s_{23}s_{13}e^{i\delta} & 
		c_{12}c_{23} - s_{12}s_{23}s_{13}e^{i\delta} & s_{23}c_{13}\cr
        s_{12}s_{23} - c_{12}c_{23}s_{13}e^{i\delta} & 
		-c_{12}s_{23} - s_{12}c_{23}s_{13}e^{i\delta} & c_{23}c_{13}\cr}
	    \right)
\label{eq:mns}
\end{equation}
in terms of $c_{12} = \cos \theta_{12},\ s_{12} = \sin \theta_{12}$, etc.
With the oscillation parameters relevant to the scenarios indicated above,
we can approximate $\theta_{13} = 0^o$ and $\theta_{23} = 45^o$ whereby 
Eq.~(\ref{eq:mns}) becomes essentially 
\begin{equation}
  U_{MNS} = \left(\matrix{c_{12} & s_{12} & 0\cr 
	-s_{12}/\sqrt{2} & c_{12}/\sqrt{2} & 1/\sqrt{2}\cr
	s_{12}/\sqrt{2} & -c_{12}/\sqrt{2} & 1/\sqrt{2}\cr}\right),
\label{eq:max}
\end{equation}
where the light neutrino mass eigenstates are given in terms of the 
flavor states by 
\begin{equation}
\begin{array}{rcl}
	\nu_3 &=& \frac{1}{\sqrt{2}}(\nu_\mu + \nu_\tau),\\[4pt]
	\nu_2 &=& \nu_e \sin \theta_{12} + \frac{1}{\sqrt{2}}
		(\nu_\mu - \nu_\tau)\cos \theta_{12},\\[4pt]
	\nu_1 &=& \nu_e \cos \theta_{12} - \frac{1}{\sqrt{2}}
		(\nu_\mu - \nu_\tau)\sin \theta_{12},\\[4pt]
\end{array}
\label{eq:states}
\end{equation}
For the SMA solution, $\theta_{12} = 1.4^o$, while the three large
mixing solar solutions differ from maximal in that the angle is approximately
$30^o$ for the LMA, $39^o$ for the LOW, and $51^o$ for the QVO solutions.
In contrast, the CKM quark mixing matrix is approximately 
\begin{equation}
	V_{CKM} = \left(\matrix{ 0.975 & 0.220 & 0.0032e^{-i\delta} \cr
		-0.220 & 0.974 & 0.040 \cr 0.0088 & - 0.040 & 0.999 \cr}\right).
\label{eq:ckm}
\end{equation}
An important issue to be answered is why $U_{\mu 3} \simeq 1/\sqrt{2}$ is so 
much larger than $V_{cb} \simeq 0.040$.

Maximal mixing of two neutrino mass eigenstates can arise if the two states
are nearly degenerate in mass, i.e., the neutrinos are pseudo-Dirac.  It 
can also arise if the determinant of the $2 \times 2$ submatrix nearly 
vanishes.  For example,
\begin{equation} 
	\left(\matrix{ x^2 & x \cr x & 1 \cr}\right) \rightarrow 
	\lambda = 0,\ 1 + x^2,\ \psi_0 \sim \left(\matrix{1 \cr -x \cr}\right),
		\ \psi_+ \sim \left( \matrix{x \cr 1 \cr}\right),
\label{eq:exam}
\end{equation}
and the components are comparable for $x \simeq 1$.  The first situation
is relevant for the QVO and LOW near maximal mixings, while the second
is more relevant for the atmospheric and LMA mixings where a sizable 
hierarchy is expected to be present.

Finally, it should be noted that the $U_{MNS}$ mixing matrix is the 
product of two unitary transformations diagonalizing the charged lepton 
mass matrix $L$ and the light neutrino mass matrix $M_\nu$:  
\begin{equation}
	U_{MNS} = U^\dagger_L U_\nu,
\label{eq:mixing}
\end{equation}
where by the seesaw mechanism $M_\nu = -N^T M^{-1}_R N$ is given in terms 
of the Dirac neutrino matrix $N$ and the right-handed Majorana matrix $M_R$.  
The structure of $U_{MNS}$ is 
then determined by the three matrices $N,\ M_R$ and $L$, one of which 
or in concert can play a role in the maximal or bimaximal mixing pattern.

\section{Types of Neutrino Models and Possible Unifications}
Neutrino models can be characterized as belonging to one of three types
for the purpose of this talk.\\[-0.2in]
\begin{itemize}
\item	Those involving only left-handed fields leading to a left-handed 
	Majorana mass matrix with no Dirac neutrino mass matrix present.  The 
	Zee model~\cite{zee} is a prime example.  Typically lepton number is 
	violated by two units, or an $L = -2$ isovector Higgs field is 
	introduced. A newly-defined lepton number $\bar{L} \equiv L_e - 
	L_\mu - L_\tau$ is often taken to be conserved.  The ultralight 
	neutrino masses, however, are not easily understood.  
\item	Models in which both left-handed and right-handed fields are 
	present.  With no Higgs contributions to the left-handed Majorana
	mass matrix, the seesaw mechanism readily yields ultralight 
	neutrino masses, provided the right-handed Majorana masses are 
	in the range of $10^5 - 10^{14}$ GeV.  Such masses are naturally
	obtained in SUSY GUT models with $\Lambda_G = 2 \times 10^{16}$
	GeV.  
\item	Models in which neutrinos probe higher dimensions.  Right-handed
	neutrinos which are singlets under all gauge symmetries can enter
	the bulk.  With large extra dimensions and the compactification 
	scale much lower than the string scale, a modified seesaw mechanism
	can generate ultralight neutrino masses.\\[-0.2in]
\end{itemize}
I shall restrict my attention in this overview to models involving both 
left-handed and right-handed neutrinos.  In this workshop, Tobe~\cite{tobe} 
addresses purely left-handed neutrino models in the context of R-parity 
violation, while Mohapatra~\cite{moh} considers models involving higher 
dimensions.

Both the nonsupersymmetric standard model (SM) and the minimum supersymmetric
extended version (MSSM) involve no right-handed neutrinos and just one
or two Higgs doublets, respectively.  Hence no renormalizable mass terms
can be constructed for the neutrinos; moreover, the renormalizable mass
terms which are present for the quarks and charged leptons have completely
arbitrary Yukawa couplings.  In order to reduce the number of free parameters
and thereby achieve some detailed predictions for the mass spectra of the 
fundamental particles, some flavor and/or family unification must be 
introduced.  This is generally done in the context of supersymmetry for 
which the desirable feature of gauge coupling unification obtains.

Flavor or vertical symmetry has generally been achieved in the framework of 
Grand Unified Theories (GUTs) which provide unified treatments of quarks
and leptons, as (some) quarks and leptons are placed in the same multiplets.
Examples involve $SU(5),\ SU(5) \times U(1),\ SO(10),\ E_6$, etc.  

The introduction of a family or horizontal symmetry, on the other hand,
enables one to build in an apparent hierarchy for masses of comparable flavors
belonging to different families.  Such a symmetry may be discrete
as in the case of $Z_2,\ S_3,\ Z_2 \times Z_2$, etc. which results in 
multiplicative quantum numbers.  A continuous symmetry such as 
$U(1),\ U(2),\ SU(3)$, etc., on the other hand, results in additive 
quantum numbers and may be global or local (and possibly anomalous).  

Combined flavor and family symmetries will typically reduce the number of
model parameters even more effectively.  On the other hand, the unification
of flavor and family symmetries into one single group such as $SO(18)$ or
$SU(8)$, for example, has generally not been successful, as too many extra 
states are present which must be made superheavy.

\section{Froggatt-Nielsen-type Models with Anomalous $U(1)$\\ Family Symmetry}
In 1979 Froggatt and Nielsen~\cite{f-n} added to the SM a scalar singlet 
``flavon'' $\phi_f$, which gets a VEV, together with heavy fermions, 
$(F,\ \bar{F})$, in vector-like representations, all of which carry 
$U(1)$ family charges.  With $U(1)$ broken at a scale $M_G$ by 
$\langle \phi \rangle/M_G \equiv \lambda
\sim (0.01 - 0.2)$, the light and heavy fermions are mixed; hence
$\lambda$ can serve as an expansion parameter for the quark and 
lepton mass matrix entries.  No GUT is involved, although $M_G$ is some
high unspecified scale.

This idea received a revival in the past decade when it was observed by 
Ibanez~\cite{ibanez} that string theories with anomalous $U(1)$'s generate 
Fayet-Iliopoulos D-terms which trigger the breaking of the $U(1)$ at a scale 
of $O(\lambda)$ below the cutoff, again providing a suitable expansion 
parameter.  The $\lambda^n$ structure of the mass matrices can be determined 
from the corresponding Wolfenstein $\lambda$ structure of the 
CKM matrix and the quark and lepton mass ratios, where different
$U(1)$ charges are assigned to each quark and lepton field.

By careful assignment of the $U(1)$ charges, Ramond and many other 
authors~\footnote{For reference listings in two more comprehensive reviews,
see \cite{u1}.} have shown that maximal mixing of $\nu_\mu \leftrightarrow 
\nu_\tau$ can be obtained, while the SMA solution for $\nu_e \leftrightarrow 
\nu_\mu,\ \nu_\tau$ is strongly favored.  Since then, other authors~\cite{alt}
have applied the technique in the presence of $SU(5)$ or $SO(10)$ to get also 
the QVO or LOW solutions.  Very recently, Kitano and Mimura~\cite{km} have 
considered $SU(5)$ and $SO(10)$ models in this framework with an $SU(3) 
\times U(1)$ horizontal symmetry to show that the LMA solution can also be 
obtained.  But with these types of models, the coefficients (prefactors) 
of the $\lambda$ powers can not be accurately predicted.

\section{Predictive SUSY GUT Models}
With the minimal $SU(5)$ SUSY GUT model extended to include the 
left-handed conjugate neutrinos, the matter fields are placed
in ${\bf \bar{5}}$ and ${\bf 10}$ representations according to 
\begin{equation}
  \bar{5}_i \supset (d^c_\alpha,\ \ell,\ \nu_\ell)_i,\quad 10_i \supset 
	(u_\alpha,\ d_\alpha,\ u^c_\alpha,\ \ell^c)_i,
	\quad 1_i \supset (\nu^c)_i,\quad \alpha = 1,2,3.  
\label{eq:5matter}
\end{equation}
while the Higgs fields are placed in the adjoint and fundamental 
representations
\begin{equation}
  \Sigma(24),\ H_u(5),\ H_d(\bar{5}).  
\label{eq:5higgs}
\end{equation}
The $SU(5)$ symmetry is broken down to the MSSM at a scale $\Lambda_G$
with $\langle \Sigma \rangle$ pointing in the $B-L$ direction, 
but doublet-triplet splitting must be done by hand.  The 
electroweak breaking occurs when the $H_u$ and $H_d$ VEV's are generated.

The number of Yukawa couplings has now been reduced in the Yukawa 
superpotential, and  the fermion mass matrices 
exhibit the symmetries, $M_U = M^T_U,\ M_D = M^T_L$.  This implies 
$m_b = m_\tau$ at the GUT scale, but also $m_d/m_s = m_e/m_\mu$  
which is too simplistic since no family symmetry is present.  One can 
circumvent this problem by introducing a family or horizontal symmetry,
but more predictive results are obtained in the $SO(10)$ framework.

In $SO(10)$ all fermions of one family are placed in a ${\bf 16}$ spinor 
supermultiplet and carry the same family charge assignment:
\begin{equation}
  {\bf 16}_i(u_\alpha,\ d_\alpha,\ u^c_\alpha,\ d^c_\alpha,\ \ell,
	\ \ell^c,\ \nu_\ell,\ \nu^c_\ell)_i,\quad i = 1,2,3.
\label{eq:10mat}
\end{equation}
Massive pairs of $({\bf 16},\ \overline{\bf 16})$'s and ${\bf 10}$'s 
may also be present. The Higgs Fields may contain one or more ${\bf 45}_H$'s 
and pairs of ${\bf 16}_H,\ \overline{\bf 16}_H$ which break $SO(10)$ down 
to the SM, while $\bf{10}_H$ breaks the electroweak group at the electroweak
scale.  A $\overline{\bf 126}_H$ or effective $\overline{\bf 16}_H\cdot
\overline{\bf 16}_H$ field can generate superheavy right-handed
Majorana neutrino masses.

With an appropriate family symmetry introduced, a number of texture zeros
will appear in the mass matrices.  These will enable one to make some 
well-defined predictions for the masses and mixings of the quarks and leptons,
for typically fewer mass matrix parameters will be present than the 20
quark and lepton mass and mixing observables plus 3 right-handed 
Majorana masses.

Yukawa $t-b-\tau$ coupling unification is possible only for 
$\tan \beta = v_u/v_d \simeq 55$ in this minimal Higgs case described above.  
However, if a 
${\bf 16'}_H,\ \overline{\bf 16'}_H$ pair is introduced with the former
getting an electroweak-breaking VEV which helps contribute to $H_d$~\cite{abb},
or if the ${\bf 16}_H$ of the first pair also gets an EW VEV, Yukawa coupling 
unification is possible for $\tan \beta \ll 55$.
Such breaking VEV's can contribute asymmetrically 
to the down quark and charged lepton mass matrices.  This makes it 
possible to understand large $\nu_\mu - \nu_\tau$ mixing, $U_{\mu 3} 
\simeq 0.707$, while $V_{cb} \simeq 0.040$.
Moreover, the Georgi-Jarlskog mass relations~\cite{gj},
\begin{equation}
	m_s/m_b = m_\mu/3m_\tau\ {\rm and}\ m_d/m_b = 3m_e/m_\tau,
\label{eq:g-j}
\end{equation}
can be generated by the mass matrices with the help of the same asymmetrical
contributions.

Models based on $SO(10)$ then differ due to their matter and Higgs contents
as well as the horizontal family symmetry group chosen.  Several selected 
illustrative examples of predictive $SO(10)$ GUT models are presented below, 
where some of their characteristic features are highlighted.

\subsection{$SO(10)$ with $U(1)_H$}
A model of this type has been presented by Babu, Pati, and Wilczek~\cite{bpw}
based on dimension-5 effective operators involving 
\begin{equation}
\begin{array}{ll}
	{\rm Matter\ Fields:} & {\bf 16}_1,\ {\bf 16}_2,\ {\bf 16}_3\\
	{\rm Higgs\ Fields:}  & {\bf 10}_H,\ {\bf 16}_H,\ \overline{\bf 16}_H,
		\ {\bf 45}_H\\
\label{eq:bpw}
\end{array}
\end{equation}
The ${\bf 16}_H$ develops both GUT and EW scale VEV's.  With no CP violation,
11 matrix input parameters yield 18 + 3 masses and mixings.  Maximal 
$\nu_\mu \leftrightarrow \nu_\tau$ mixing arises from the seesaw mechanism,
while the SMA solar solution is preferred.

\subsection{$SO(10)$ with $\left[U(1) \times Z_2 \times Z_2\right]_H$}
Barr and Raby~\cite{br} have shown that a stable solution to the 
doublet-triplet splitting problem in $SO(10)$ can be obtained based on this 
global horizontal group.  With an extension of the minimal Higgs 
content involved, Albright and Barr~\cite{ab} have developed a model
involving only renormalizable terms in the Yukawa superpotential with the 
following superfields present:
\begin{equation}
\begin{array}{ll}
  {\rm Matter\ Fields:} & {\bf 16}_1,\ {\bf 16}_2,\ {\bf 16}_3,\ 
	2({\bf 16},\ \overline{\bf 16})'s,\ 2({\bf 10})'s,\ 6({\bf 1})'s\\
  {\rm Higgs\ Fields:} & 4({\bf 10}_H)'s,\ 2({\bf 16}_H,\ \overline{\bf 16}_H)
	's,\ {\bf 45}_H,\ 5({\bf 1}_H)'s\\
\label{eq:ab}
\end{array}
\end{equation}
Ten matrix input parameters yield all 20 + 3 masses and mixings.  A value
of $\tan \beta \simeq 5$ is favored with $\sin 2\beta \sim 0.65$ obtained
for the CKM unitarity triangle.  Maximal $\nu_\mu \leftrightarrow \nu_\tau$
mixing arises from the lopsided texture of the charged lepton matrix.  In 
this simplest scenario, the QVO solution is preferred, although by modifying
the right-handed Majorana matrix the SMA or LMA solar solutions can be 
obtained with one or four more input parameters, respectively.

\subsection{$SO(10)$ with $\left[SU(2) \times Z_2 \times Z_2 \times 
	Z_2\right]_H$}
Chen and Mahanthappa~\cite{cm} have based a model on this family group with
dim-5 effective operators involving the following superfields:
\begin{equation}
\begin{array}{ll}
  {\rm Matter\ Fields:} & ({\bf 16},2),\ ({\bf 16},1)\\
  {\rm Higgs\ Fields:} & 5({\bf 10},1)_H{'s},\ 3(\overline{\bf 126},1)_H{'s}\\
  {\rm Flavon\ Fields:} & 3({\bf 1},2)_H{'s},\ 3({\bf 1},3)_H{'s}\\
\label{eq:cm}
\end{array}
\end{equation}
With no CP violation, eleven matrix input parameters yield 18 + 3 masses 
and mixings, while $\tan \beta = 10$ is assumed.  Maximal $\nu_\mu 
\leftrightarrow \nu_\tau$ mixing arises from the seesaw mechanism with symmetric
mass matrices.  The QVO solar solution is preferred, while the SMA solution 
or the LMA solution with $\tan^2 \theta_{sol} < 1$ is difficult to obtain.

\subsection{$SO(10)$ with $\left[U(2) \times U(1)^n\right]_H$}
Blazek, Raby, and Tobe~\cite{brt} have constructed such a model involving only 
renormalizable terms in the Yukawa superpotential with the following fields:
\begin{equation}
\begin{array}{ll}
  {\rm Matter\ Fields:} & ({\bf 16},2),\ ({\bf 16},1),\ ({\bf 1},2),\ 
	({\bf 1},1)\\
  {\rm Higgs Fields:} & ({\bf 10},1)_H,\ ({\bf 45},1)_H\\
  {\rm Flavon\ Fields:} & 2({\bf 1},2)_H{'s},\ ({\bf 1},3)_H,\ 
	2({\bf 1},1)_H{'s}\\
\label{eq:brt}
\end{array}
\end{equation}
Sixteen matrix input parameters yield the 20 + 3 masses and mixings with 
$\tan \beta \simeq 55$ required.  CP violation occurs with $\sin 2\beta$ in
the second quadrant for the CKM unitarity triangle.  All solar neutrino
solutions, SMA, LMA, LOW and QVO, are possible.

\subsection{$SO(10)$ with $\left[SU(3) \times {\rm unspecified\ discrete\ 
	symmetries}\right]_H$}
Berezhiani and Rossi~\cite{brossi} have proposed a model based on this 
group with the following structure:
\begin{equation}
\begin{array}{ll}
  {\rm Matter\ Fields:} & ({\bf 16},3),\ ({\bf 16},\overline{3}),\ 
	(\overline{\bf 16},3),\ 2({\bf 16},3)'s,\\
	& 2(\overline{\bf 16},\overline{3})'s,\ ({\bf 1},\overline{3}),\ 
	  ({\bf 1},3),\ ({\bf 10},3),\ ({\bf 10},\overline{3})\\
  {\rm Higgs\ Fields:} & ({\bf 16},1)_H,\ (\overline{\bf 16},1)_H,\ 
	({\bf 54},1)_H,\ 2({\bf 45},1)_H{'s},\\
	& 2({\bf 10},1)_H{'s}\\
  {\rm Flavon\ Fields:} & ({\bf 1},\overline{6})_H,\ 3({\bf 1},3)_H{'s},\ 
	({\bf 1},8)_H\\
\label{eq:brossi}
\end{array}
\end{equation}
Fourteen matrix input parameters yield 18 + 3 masses and mixings with
moderate $\tan \beta$ assumed.  Maximal $\nu_\mu \leftrightarrow \nu_\tau$
mixing arises from the lopsided texture of the charged lepton mass matrix.
The SMA solar solution is preferred, while other solutions are possible 
with modification of the right-handed Majorana matrix.

\section{Concluding Remarks}
The most predictive models for the 12 ``light'' fermion masses and their
8 CKM and MNS mixing angles and phases are obtained in the framework of 
grand unified models with family symmetries.  The $SO(10)$ models are more
tightly constrained than $SU(5)$ models and are more economical than
larger groups such as $E_6$~\cite{ma}, where more fields must be made 
supermassive. In fact, some $SO(10)$ models do very well in predicting the 
20 + 3 ``observables'' with just 10 or more input parameters, depending on the 
model and type of solar neutrino mixing solution involved.

The SMA MSW solution is readily obtained in many unified models, since 
only one pair of states, $\nu_\mu$ and $\nu_\tau$, are maximally mixed.
Bimaximal mixing can be obtained in a smaller class of models, with the 
QVO solution having the more natural hierarchy with a pair of pseudo-Dirac
neutrinos.  The LMA MSW solution, if allowed, requires the most fine tuning,
for two nearly maximal mixings must be obtained with a hierarchy of 
neutrino masses.  For this latter solution, the right-handed Majorana
neutrino masses typically span a range of $10^6 - 10^{14}$ GeV, while the 
lightest is typically $10^{10}$ GeV for the other solar neutrino solutions.

Unfortunately, while the experimental solar neutrino solution remains
rather uncertain, unified model builders are not able to clarify the 
situation by predicting the outcome with any degree of certainty.

\end{document}